\documentclass[10pt,a4paper]{article}
\usepackage{amsmath,latexsym,amssymb,amsfonts}
\usepackage[dvips]{graphicx}
\usepackage{bm}

\begin{document}

\title{\textbf{No-boundary wave function for two-field inflation}}
\author{\textsc{Dong-il Hwang}$^{a}$\footnote{{\tt dongil.j.hwang@gmail.com}},\;\; \textsc{Soo A Kim}$^{b}$\footnote{{\tt sooastar@googlemail.com}}\;\; and \textsc{Dong-han Yeom}$^{c,d}$\footnote{{\tt innocent.yeom@gmail.com}}\\
\textit{$^{a}$\small{Institute of Basic Science, Sogang University, Seoul 121-742, Republic of Korea}}\\
\textit{$^{b}$\small{Institute for the Early Universe, Ewha Womans University, Seoul 120-750, Republic of Korea}}\\
\textit{$^{c}$\small{Yukawa Institute for Theoretical Physics, Kyoto University, Kyoto 606-8502, Japan}}\\
\textit{$^{d}$\small{Leung Center for Cosmology and Particle Astrophysics, National Taiwan University, Taipei 10617, Taiwan}}
}
\maketitle

\begin{abstract}
In this paper, we investigate the no-boundary wave function and the complex-valued instantons for two-field inflation models that have different masses. If there is a relatively massive direction, to classicalize the massive field, the solution should start from the slow direction with relatively larger vacuum energy. Therefore, the existence of the massive direction implies the increase of expected $e$-foldings. The most probable $e$-foldings are approximately $\mathcal{N} \simeq (m_{2}/m_{1})^{2} \times \mathcal{O}(1)$ in the $m_{1} \ll m_{2}$ limit. Therefore, as long as there is a sufficient mass hierarchy, the no-boundary wave function can reasonably explain large $e$-foldings, so to speak more than $50$ $e$-foldings.
\end{abstract}

\begin{flushright}
{\tt YITP-14-24}
\end{flushright}

\newpage
\tableofcontents
\newpage

\section{Introduction}

Understanding the beginning of our universe is the important task of modern physics and cosmology. The theory of quantum gravity and the application to cosmology should resolve the problem of the initial singularity \cite{Hawking:1969sw} and also should give a reasonable probability to explain the initial conditions for our universe, especially the initial conditions for inflation \cite{Guth:1980zm}. Now, we are getting data from cosmological observations and soon after we will be able to understand the detailed mechanism for inflation. Perhaps, the recent tension between the Planck data \cite{Ade:2013uln} and the BICEP2 results \cite{bicep2} may require multi-field inflation or complication of a single field inflation model, though it is not possible to conclude yet. In this context, now this is a natural question: \textit{can the multi-field inflation reasonable to explain our inflationary universe in terms of quantum gravity?}

Following the canonical quantization \cite{DeWitt:1967yk}, the master wave function that contains all information of our universe, so-called the wave function of the universe, is governed by the Wheeler-DeWitt equation. The solution depends on the boundary condition. Perhaps one natural assumption is the ground state, where the ground state wave function can be obtained by the Euclidean path integral \cite{Hartle:1983ai} (there can be alternative boundary conditions, e.g., \cite{Vilenkin:1986cy}). This wave function is known as the \textit{no-boundary wave function} and this is called by the no-boundary proposal. The Euclidean path integral is approximated by sum-over on-shell solutions, so-called instantons\footnote{Since we are using the instanton method, in other words a kind of semi-classical methods, this approach may lose a truly wave nature of the entire wave function, e.g., a resonant structure \cite{Maydanyuk:2010qb}. However, as long as the quantum state is in the ground state and the parameters allow a regime where the steepest-descent approximation is still sound, the instanton approximations will be a good description that describes the no-boundary wave function.}. In general, these instantons are complex-valued \cite{Halliwell:1989dy}. After a long Lorentzian time, the instanton should return to real-valued functions \cite{Lyons:1992ua,Hartle:2007gi,Hartle:2008ng}. This condition is called by the \textit{classicality}; note that in this paper we use the terminology `classicality' for the classicalization of all matter fields and the metric, while sometimes the same terminology is used to explain the classicalization of inhomogeneous perturbations generated during inflation.

To impose the classicality, each history needs a period of slow-roll inflation \cite{Hartle:2008ng}; and the boundary of the classicalizable region forms a \textit{cutoff} near the local minimum such that if the initial condition is inside the cutoff, the history cannot be classicalized. However, one traditional problem of the no-boundary wave function (with Einstein gravity and single field inflation) is that the result does not prefer large $e$-foldings. For classicality, we only need just order one $e$-foldings. To explain large $e$-foldings, e.g., more than $50$ $e$-foldings, we need further additional assumptions. For example, Hartle, Hawking and Hertog weighted the volume factor to enhance the initial conditions for large $e$-foldings \cite{Hartle:2007gi,Hawking:2002af}. Apart from this \textit{ad hoc} assumption, one may introduce other reasonable assumptions that can enhance large $e$-foldings \cite{Hwang:2013nja}, e.g., introducing a (perhaps, Planck scale) pre-inflation era before the primordial inflation of our universe, finely tune the shape of the potential, introducing the multi-field inflation scenario \cite{Hwang:2012bd}, or using new contributions that can come from modified gravity \cite{Sasaki:2013nka}.

However, these previous analysis in \cite{Hwang:2013nja} relied on the single field inflation. Even for the multi-field inflation case, we only analyzed for the single mass case that is effectively equivalent to single field inflation. On the other hand, more realistic inflation model will be cooperated by the contributions of many fields with various potential shapes. In this paper, to investigate this issue, as a toy model, we study classicalized instantons for two-field inflation models with different mass parameters. We observe that the cutoff structure drastically changes and hence we have to change the naive intuitions that come from the result of the single field inflation.

In Section~\ref{sec:int}, we describe the formulation of the no-boundary proposal for the two-field inflation model. In Section~\ref{sec:cla}, we discuss the motivations of this paper. In Section~\ref{sec:two}, we study the no-boundary wave function for the two-field inflation model by using analytic and numerical methods. Finally, in Section~\ref{sec:con}, we summarize our conclusions.

\section{\label{sec:int}No-boundary wave function for two scalar fields}

The ground state wave function that was suggested by Hartle and Hawking \cite{Hartle:1983ai} is defined as the Euclidean path-integral on a compact $3$-dimensional manifold $\Sigma$ as a function of $3$-metric $h_{\mu\nu}$ and a field value $\chi$ by
\begin{eqnarray}
\Psi[h_{\mu\nu}, \chi] = \int_{\mathcal{M}} \mathcal{D}g_{\mu\nu} \mathcal{D} \Phi \; e^{-S_{\text{E}}[g_{\mu\nu}, \Phi]},
\label{general_wave_function}
\end{eqnarray}
where the $4$-metric $g_{\mu\nu}$ and the field $\Phi$ (for multi-field case, include all fields) take the value $h_{\mu\nu}$ and $\chi$ on $\Sigma$. Here, we sum-over all compact $4$-dimensional Euclidean manifolds that have $\Sigma$ as their only boundary.

In this paper, we investigate Einstein gravity with two minimally coupled scalar fields $\Phi_{1,2}$ (we choose the units $c=G= \hbar = 1$):
\begin{eqnarray}
S_{\text{E}} = - \int d^{4}x \sqrt{+g} \left( \frac{1}{16\pi} R - \sum_{i=1,2} \frac{1}{2} (\nabla \Phi_{i})^{2} - V(\Phi_{1},\Phi_{2}) \right).
\end{eqnarray}
We consider the scalar fields with the quadratic potentials with mass $m_{1}$ and $m_{2}$:
\begin{eqnarray}
V(\Phi_{1},\Phi_{2}) = \frac{1}{2} m_{1}^{2} \Phi_{1}^{2} + \frac{1}{2} m_{2}^{2} \Phi_{2}^{2}.
\end{eqnarray}

\paragraph{Minisuperspace model}

We impose the minisuperspace model following the $O(4)$ symmetric metric ansatz
\begin{eqnarray}
ds^{2} = m_{2}^{-2} \left[N(\lambda)^{2} d\lambda^{2} + a(\lambda)^{2} d\Omega_{3}^{2}\right].
\end{eqnarray}
From this choice of metric, it is convenient to redefine the field by
\begin{gather}
\phi_{i} \equiv \sqrt{\frac{4 \pi}{3}} \Phi_{i},
\end{gather}
though later we will use the notation of $\Phi_{i}$ again in some places. The no-boundary wave function is now
\begin{eqnarray}
\Psi[b,\chi_{1},\chi_{2}] = \int_{\mathcal{C}} \mathcal{D}N \mathcal{D}a \mathcal{D}\phi_{1} \mathcal{D}\phi_{2} \; e^{-S_{\text{E}}[a,\phi_{1},\phi_{2}]},
\label{minisuperspace_wave_function}
\end{eqnarray}
where the action is reduced by
\begin{eqnarray}
S_{\text{E}} = \frac{3 \pi}{4m_{2}^{2}} \int N d\lambda \left\{- a \left(\frac{d a}{N d\lambda}\right)^{2} - a + a^{3}\left[\left(\frac{d \phi_{1}}{N d\lambda}\right)^{2} + \left(\frac{d \phi_{2}}{N d\lambda}\right)^{2} + \frac{m_{1}^{2}}{m_{2}^{2}}\phi_{1}^{2} +  \phi_{2}^{2} \right] \right\}.
\end{eqnarray}
Along the contour $\mathcal{C}$, the metric $a$ starts in the Euclidean signature from zero, which is called by the South Pole.
It grows to the boundary value $b$ in the Lorentzian regime where $\phi_{i}$ takes the value $\chi_{i}$. Note that the action can be scaled by $m_{2}$ and hence the dynamics only depends on the mass ratio $m_{1}/m_{2}$. Therefore, for numerical calculations, we first fix $m_{2} = 1$ and only vary $m_{1}/m_{2}$ without loss of generality.

\paragraph{Steepest-descent approximation}

To calculate the path-integral, we further use the steepest-descent approximation, and hence we approximate the wave function by sum-over on-shell paths satisfying the boundary condition, so-called \emph{instantons}. For such paths $p$ which extremize the action, the no-boundary wave function is approximated to
\begin{eqnarray}
\Psi[b,\chi_{1},\chi_{2}] \simeq \sum_{p} e^{-S_{\text{E}}^{p}[b,\chi_{1},\chi_{2}]}.
\end{eqnarray}
We define a parameter $\tau$ to specify the integration contour,
\begin{eqnarray}
\tau(\lambda) \equiv \int^{\lambda} d\lambda' N(\lambda').
\label{integration_contour}
\end{eqnarray}
Then, the on-shell Euclidean action and equations of motion are
\begin{eqnarray}
0 &=& \ddot{a} + a \left(2 \dot{\phi}_{1}^{2} + 2 \dot{\phi}_{2}^{2} + \left(\frac{m_{1}^{2}}{m_{2}^{2}}\right)\phi_{1}^{2} +  \phi_{2}^{2} \right),\\
0 &=& \ddot{\phi}_{1} + 3 \frac{\dot{a}}{a} \dot{\phi}_{1} - \left(\frac{m_{1}^{2}}{m_{2}^{2}}\right) \phi_{1},\\
0 &=& \ddot{\phi}_{2} + 3 \frac{\dot{a}}{a} \dot{\phi}_{2} - \phi_{2},
\label{equations_of_motion}
\end{eqnarray}
where $\dot{~}$ denotes a derivative with respect to $\tau$.

\paragraph{Classicality condition}

We can rearrange the action by using the DeWitt metric $G_{AB}$ such that
\begin{eqnarray}
S_{\text{E}} = \frac{3 \pi}{2} \int N d\lambda \left[\frac{1}{2} G_{AB} \left(\frac{d q^{A}}{N d\lambda}\right) \left(\frac{d q^{B}}{N d\lambda}\right) + \mathcal{V}(q^{A})\right],
\end{eqnarray}
where the canonical variables $q^{A}= \left(a \quad \phi_{1} \quad \phi_{2} \right)$ are the directions of the field space. The DeWitt metric $G_{AB}$ and the superspace potential $\mathcal{V}(q^{A})$ are defined by
\begin{gather}
G_{AB} = \left(
  \begin{array}{ccc}
    -a & 0 & 0\\
    0 & a^{3} & 0 \\
    0 & 0 & a^{3} \\
  \end{array}
\right),\nonumber\\
\mathcal{V}(q^{A}) = \frac{1}{2}\left(-a + a^{3} \left(\frac{m_{1}^{2}}{m_{2}^{2}}\right) \phi_{1}^{2} + a^{3} \phi_{2}^{2}\right).
\end{gather}
The lapse function $N$ ensures the invariance of system under the reparametrizations of parameter $\lambda$. It leads to a constraint on the Hamiltonian,
\begin{eqnarray}
H(p_{A}, q^{B}) = \frac{1}{2} G^{AB} p_{A} p_{B} + \mathcal{V}(q^{A}) = 0,
\end{eqnarray}
where $p^{A}$ is the conjugate momentum and $G^{AB}$ is the inverse DeWitt metric.

Canonical quantization is implemented by applying above classical constraint to the wave function,
\begin{eqnarray}
H\left(-i\hbar \frac{\partial}{\partial q^{A}}, q^{B}\right) \Psi(q^{A})  = \left(-\frac{\hbar^{2}}{2}\nabla^{2} + \mathcal{V}(q^{A})\right) \Psi(q^{A}) = 0,
\label{Wheeler_DeWitt}
\end{eqnarray}
where the Laplace operator is defined on the superspace. This is the Wheeler-DeWitt equation in this minisuperspace.

Since our universe has Lorentizian signature, the path integral in Equation~\eqref{minisuperspace_wave_function} should connect the Euclidean manifold to the Lorentzian manifold. Therefore, the integration contour $\tau$ is defined on the complex plane.
Although the boundary value of the scale factor and scalar fields, $b$ and $\chi_{i}$, are real valued, they are naturally complexified along this complex contour. The Euclidean action also becomes complex.

The equations of motion can be derived from the action using the Hamilton-Jacobi equation. If the action along a history is complex-valued and rapidly varies for both of real and imaginary sectors, then the history is not classical. However, if the real part of the Euclidean action varies slowly compared to the imaginary part which corresponds to the Lorentzian action, then the classical Hamilton-Jacobi equation will be approximately recovered. It is called by the classicality condition:
\begin{eqnarray}
\left|\nabla_{A} S_{\text{E}}^{\text{Re}}[b,\chi_{1},\chi_{2}]\right| \ll \left|\nabla_{A} S_{\text{E}}^{\text{Im}}[b,\chi_{1},\chi_{2}]\right|,
\label{classicality_condition}
\end{eqnarray}
where $A=b,\chi_{i}$. Throughout this paper, we will denote the real and the imaginary part by superscripts $\text{Re}$ and $\text{Im}$, respectively. When the classicality condition is satisfied, we can interpret that a classical universe arises from the quantum theory.

The classicality condition can be quantified by defining the classicality ratio. It is the ratio between the variation of real and imaginary part of the action in each direction of canonical coordinate,
\begin{eqnarray}
Cl_{A} \equiv \frac{\left|\nabla_{A} S_{\text{E}}^{\text{Re}}[b,\chi_{1},\chi_{2}]\right|}{\left|\nabla_{A} S_{\text{E}}^{\text{Im}}[b,\chi_{1},\chi_{2}]\right|}.
\end{eqnarray}
For a classical history, the classicality ratio in all field directions should become small in the late Lorentzian regime.

Now, for this classical universe, one can approximate the probability of the Wheeler-DeWitt equation by
\begin{eqnarray}
P[b,\chi_{1},\chi_{2}] \varpropto \left|\Psi[b,\chi_{1},\chi_{2}]\right|^{2} \simeq e^{-2 S_{\text{E}}^{\text{Re}}[b,\chi_{1},\chi_{2}]}
\label{no-boundary_measure}
\end{eqnarray}
which is defined on a spacelike hypersurface. Since $S_{\text{E}}^{\text{Re}}$ is approximately constant when the classicality condition is satisfied, it gives a well-defined probability measure for the ensemble of classical universes.

\begin{figure}
\begin{center}
\includegraphics[scale=0.225]{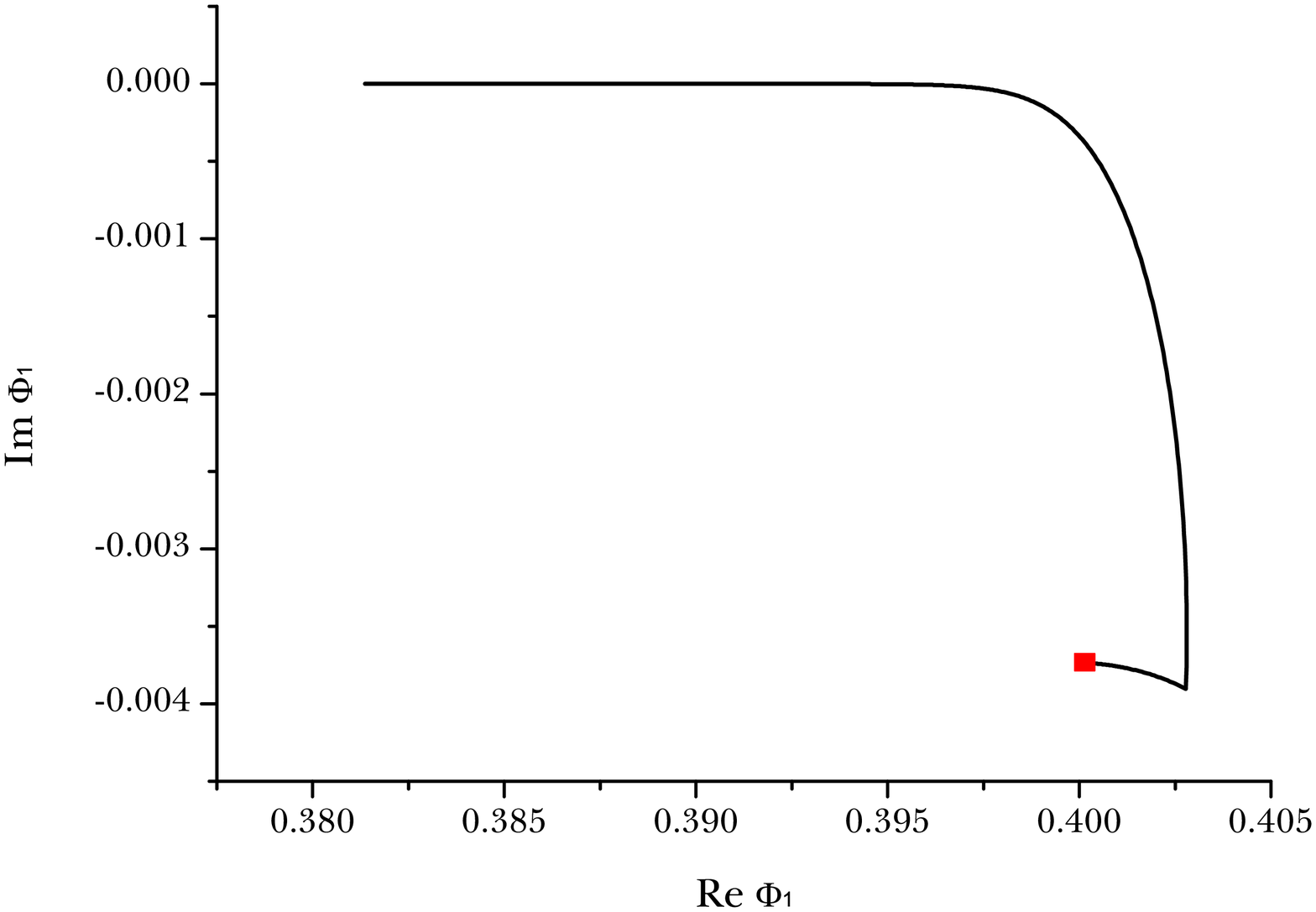}
\includegraphics[scale=0.225]{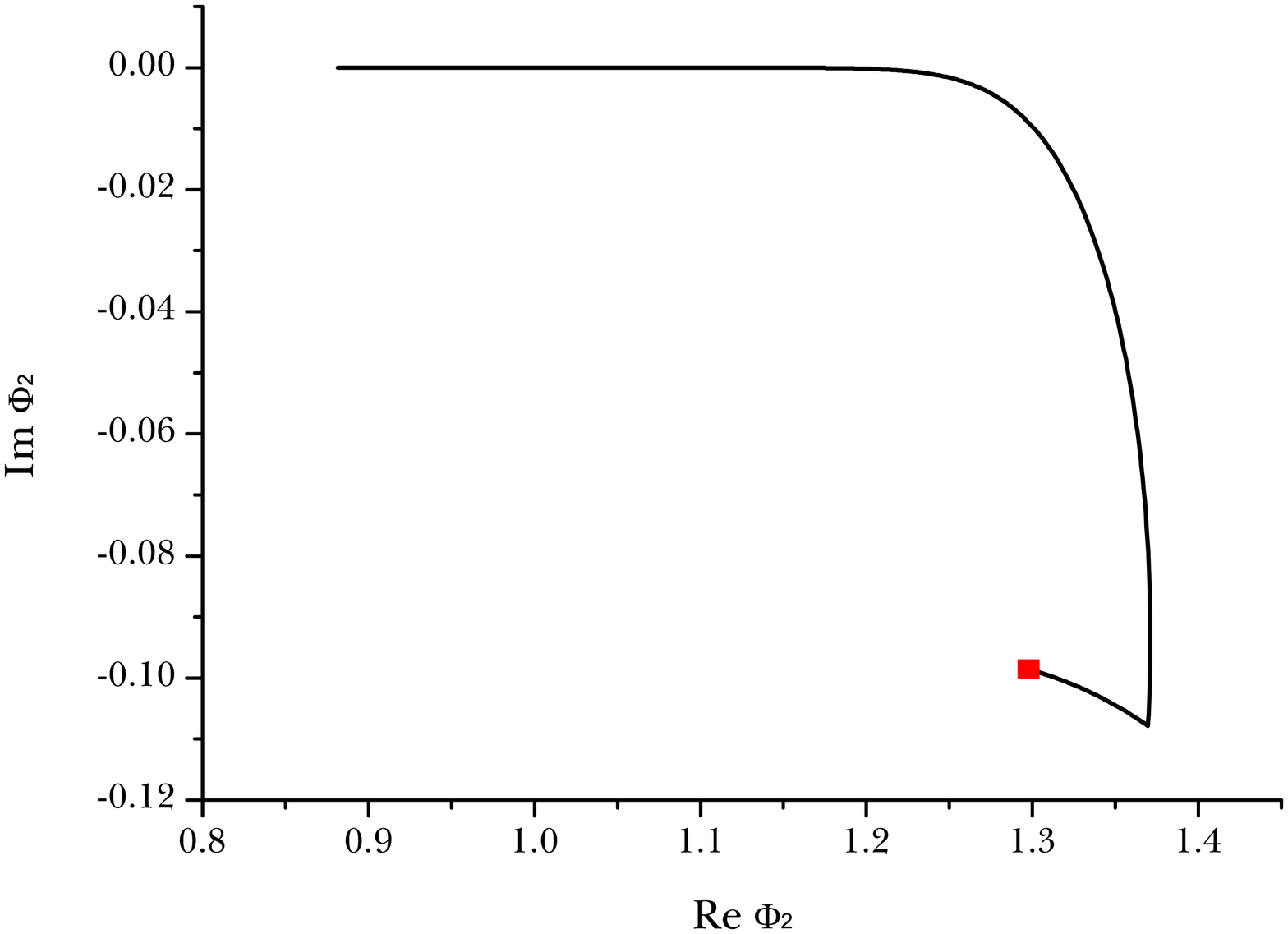}
\includegraphics[scale=0.225]{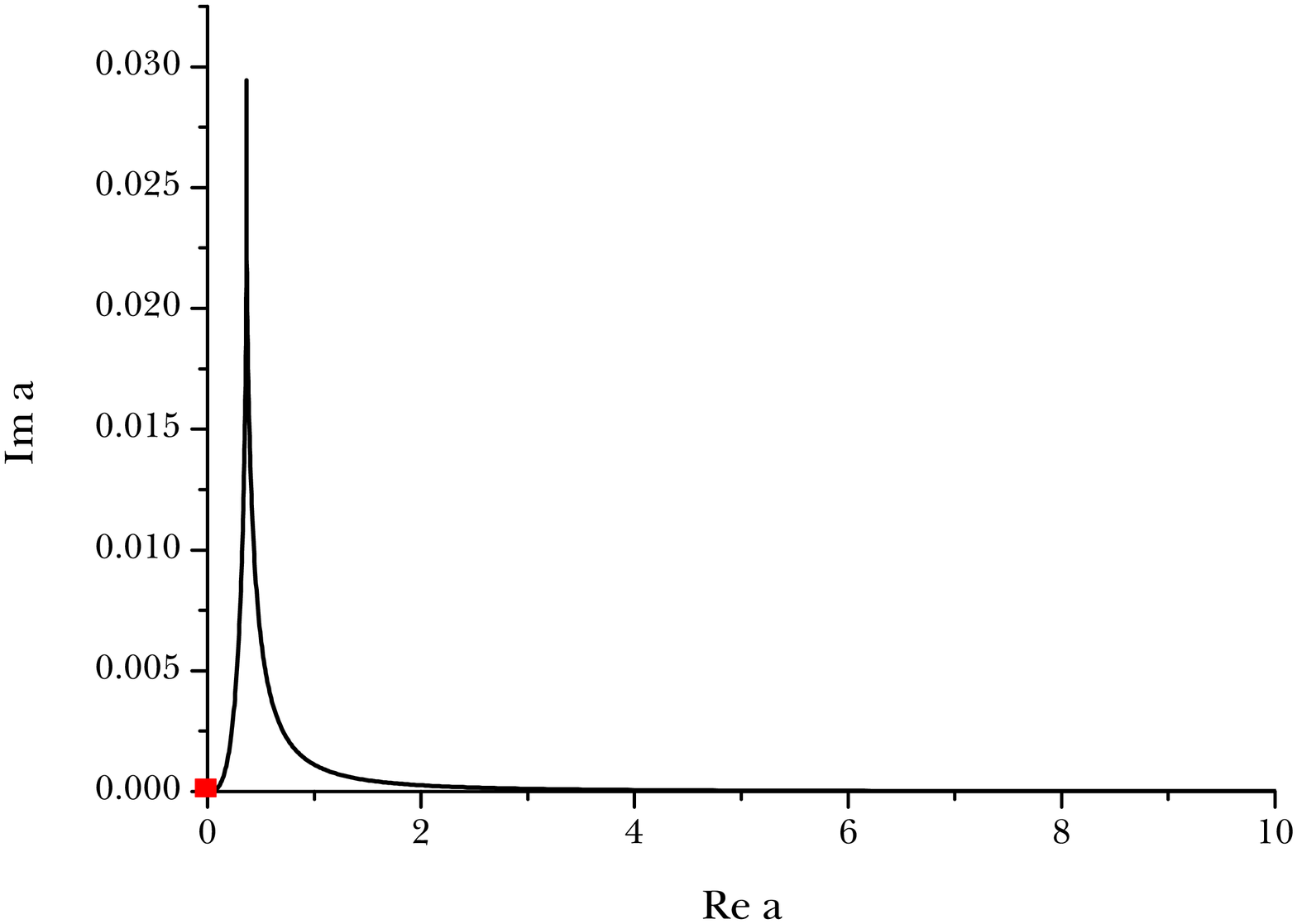}
\includegraphics[scale=0.225]{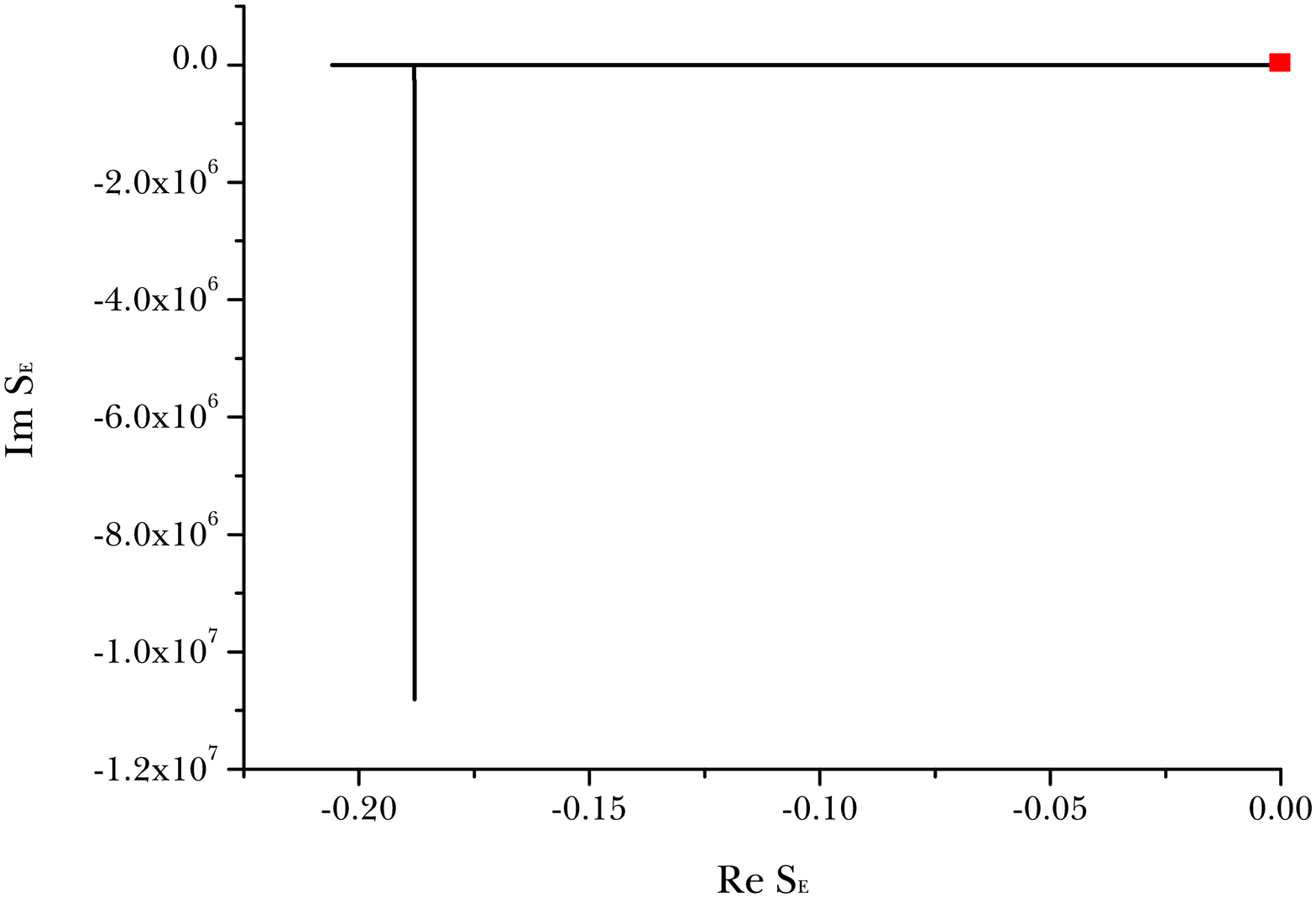}
\caption{\label{fig:ex}Plots of $\Phi_{1}$, $\Phi_{2}$, $a$, and $S_{\mathrm{E}}$ as one example of classicalized solutions, for $(m_{1}/m_{2})^{2} = 0.125$, $|\Phi_{1}(0)|=0.4$, and $|\Phi_{2}(0)|=1.3$. The red squares are the starting point.}
\end{center}
\end{figure}

\paragraph{Choosing the integration contour}

Now we choose the integration contour in Equation~\eqref{integration_contour} which connects the South Pole to the boundary where the wave function is defined. The boundary values depend only on the endpoints for smooth transformation of the contour. Therefore, we are free to choose a simple contour $\tau = x + i y$ for $0 \leq x \leq X$ and $0 \leq y \leq Y$. It connects the South Pole at $\tau = 0$ to the turning point at $\tau = X$ and this part belongs to the Euclidean regime. Then, the Lorentzian regime follows to the boundary at $\tau = X + i Y$.

Now we need to fix the boundary conditions to complete the boundary value problem.
The boundary condition at the South Pole comes from the regularity condition,
\begin{gather}
a(\tau = 0) = 0,\qquad
\dot{a}(\tau = 0) = 1,\qquad
\dot{\phi}_{i}(\tau = 0) = 0.
\end{gather}
At the end endpoint, the no-boundary wave function is imposed to have arguments $b$ and $\chi_{i}$,
\begin{gather}
a(\tau = X + i Y) = b,\qquad
\phi_{i}(\tau = X + i Y) = \chi_{i}.
\end{gather}
At the turning time, the velocities should satisfy the Cauchy-Riemann condition for the analyticity,
\begin{gather}
\frac{\partial a}{\partial x}(\tau = X) = \frac{\partial a}{i \partial y}(\tau = X),\qquad
\frac{\partial \phi_{i}}{\partial x}(\tau = X) = \frac{\partial \phi_{i}}{i \partial y}(\tau = X).
\end{gather}

The problem defined above is governed by second order differential equations of three complex functions: $a$ and $\phi_{i}$.
We have eight boundary conditions at the South Pole and three conditions at the end endpoint. Note that we further impose the classicality condition which restricts the rate of change of the action. We solve this problem by choosing a scalar field value at the South Pole,
\begin{align}
\phi_{i}(\tau = 0) \equiv \phi_{i}(0) = |\phi_{i}(0)| e^{i \theta_{i}},
\end{align}
where $|\phi_{i}(0)|$ and $\theta_{i}$ are real. Then, an initial value problem is defined so that one can evolve $a$ and $\phi_{i}$ from the South Pole. For a suitable choice of $\phi_{i}(0)$ and the turning time $\tau=X$, the classicality conditions are satisfied for the Lorentzian domain. Then, we are free to choose the end point $\tau=Y$ and take the boundary value by $b=a(\tau=X + i Y)$ and $\chi_{i}=\phi_{i}(\tau=X + i Y)$. For a given $|\phi_{i}(0)|$, one can finely tune $X$ and $\theta_{i}$ to find a solution which satisfies the classicality conditions. To do this, we used a numerical searching algorithm that was already used in \cite{Hwang:2011mp} (see Appendix A). Figure~\ref{fig:ex} is an example of the classicalizeid complex-valued instantons. One can easily see that all imaginary part approaches zero and eventually the real part of the Euclidean action is invariant.

\begin{figure}
\begin{center}
\includegraphics[scale=0.75]{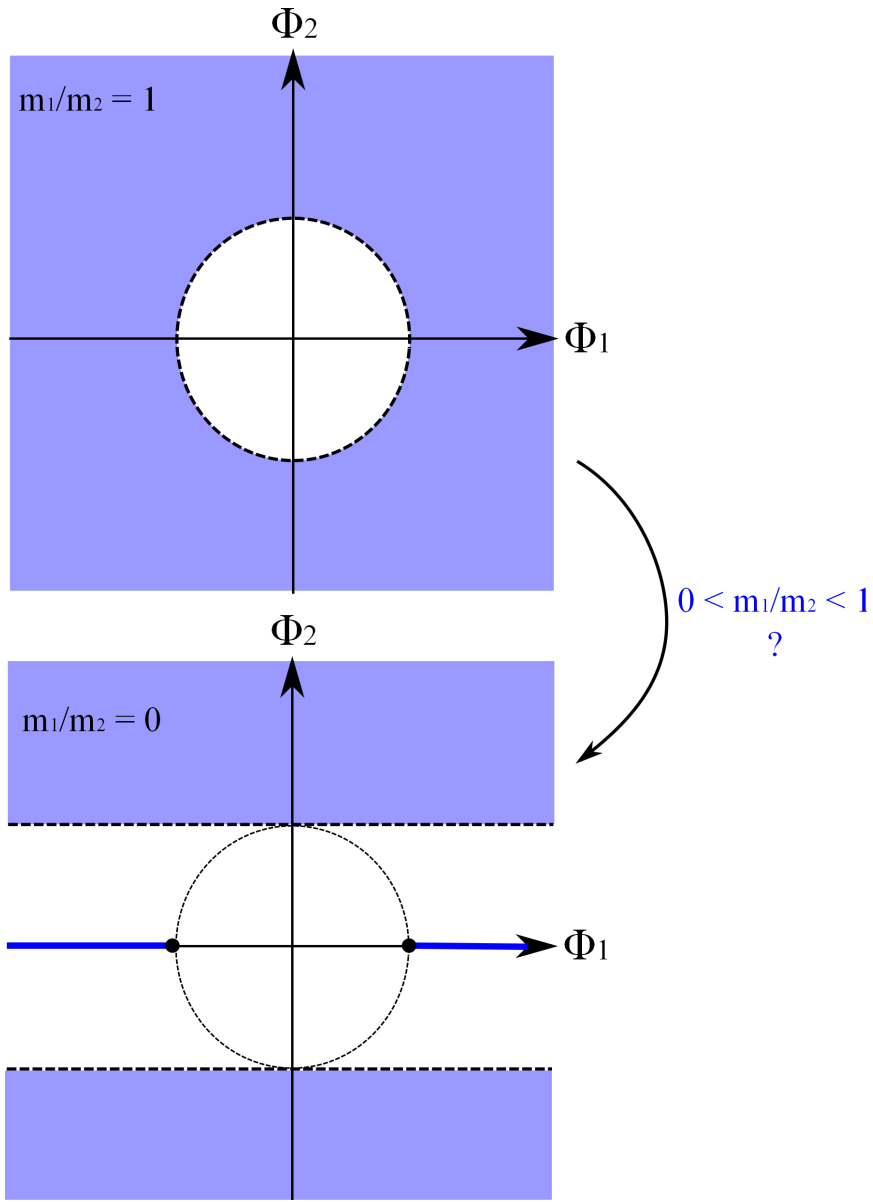}
\caption{\label{fig:concept2}The cutoff structures for $m_{1}/m_{2}=1$ (upper) and $m_{1}/m_{2} = 0$ (lower) limits. Blue colored regions are the classicalized region.}
\end{center}
\end{figure}

\section{\label{sec:cla}Motivation: cutoff around the local minimum}

Using the previous framework, we can write the probability distribution using two dimensional field space, i.e., using $|\Phi_{1}(0)|$ and $|\Phi_{2}(0)|$. In addition, since there is a symmetry between $\Phi_{1} \rightarrow - \Phi_{1}$ and $\Phi_{2} \rightarrow - \Phi_{2}$, there are four solutions for a given set of $|\Phi_{1}(0)|$ and $|\Phi_{2}(0)|$. For convenience, when we denote the Euclidean action distribution or the probability, we abuse the notation such that $P[\Phi_{1},\Phi_{2}]$ where $\Phi_{i}$ denotes $|\Phi_{i}(0)|$ if $-\pi/2 \leq \theta_{i} < \pi/2$ and $-|\Phi_{i}(0)|$ if $\pi/2 \leq \theta_{i} < 3\pi/2$. Then we can write the initial condition space $(\Phi_{1},\Phi_{2})$, where $-\infty < \Phi_{1,2} < \infty$. For a given $(\Phi_{1},\Phi_{2})$, it corresponds a unique classicalized instanton.

\paragraph{Single field model} First, let us summarize the behavior of the single field inflation model. For a given cosmological constant $\Lambda$ and the scalar field mass $m$, if $\mu \equiv m\sqrt{3/\Lambda} < 3/2$, we can find classical instantons for all range of $\Phi$. However, a qualitative difference arises when $\mu > 3/2$ which is the case of our universe \cite{Hartle:2008ng,Hwang:2012mf} (for analytic discussion, see Appendix B). In this regime, there is a critical value $\Phi_{c}$ so that for $\Phi < \Phi_{c}$, there is no classical solution for any choice of $X$ and $\theta$. In this paper, we consider the case that $\Lambda = 0$.

\paragraph{The case for $m_{1}/m_{2} = 1$:} If there are two scalar fields with the same mass $m_{1} = m_{2}$ without the cosmological constant, then the cutoff should be a circle such that $\Phi_{c}^{2} = \Phi_{1}^{2} + \Phi_{2}^{2}$ (upper of Figure~\ref{fig:concept2}). For the same mass case, the behavior can be easily generalized to the multi-field case \cite{Hwang:2012bd}.

\paragraph{The case for $m_{1}/m_{2} = 0$:} For an extreme limit, if there is one massless direction $m_{1} = 0$ and one massive field $m_{2}$, then $|\Phi_{2}| = \Phi_{c}$ is the cutoff. If $m_{1} = \epsilon \ll m_{2}$, then still approximately $|\Phi_{2}| > \Phi_{c}$ will be the classicalized region (blue colored region in lower of Figure~\ref{fig:concept2}). One interesting point is that, anyway if $m_{1} = \epsilon$ is not zero, then along the $\Phi_{2} = 0$ slice, the classical histories should appear for $|\Phi_{1}| > \Phi_{c}$ (blue thick lines in lower of Figure~\ref{fig:concept2}).

\paragraph{Motivation of this paper} Let us imagine this situation. If we vary the mass ratio $m_{1}/m_{2}$ from $1$ to $0$, then how the cutoff structure will be changed? Initially, it should begin from the circular shape. And, in the end, as the mass ratio decreases, there should appear two \textit{separated} regions, where one is \textit{two dimensional} area $|\Phi_{2}| > \Phi_{c}$ and the other is the \textit{one dimensional} slice $\Phi_{2} = 0$ and $|\Phi_{1}| > \Phi_{c}$. However, how can these two shapes of cutoffs be smoothly connected? This is the task of this paper.

\section{\label{sec:two}Classicalization of two-field inflation}

In this section, we report theoretical and numerical analyses of two field inflation models. As we vary the mass ratio between two fields, we observe the cutoffs and the probability distribution.

\subsection{Theoretical considerations}

Let us consider a situation that there are two scalar fields with the potential
\begin{eqnarray}
V(\Phi_{1},\Phi_{2}) = \frac{1}{2} m_{1}^{2} \Phi_{1}^{2} + \frac{1}{2} m_{2}^{2} \Phi_{2}^{2}. \nonumber
\end{eqnarray}
If one field $\Phi_{1}$ slowly roles, then one can approximate such that
\begin{eqnarray}
V(\Phi_{2}) \simeq V_{0} + \frac{1}{2} m_{2}^{2} \Phi_{2}^{2}
\end{eqnarray}
with $V_{0} = (1/2) m_{1}^{2} \Phi_{1}^{2}$. Then we can define the effective mass for the $\Phi_{2}$ direction (following Equation~(\ref{eq:mu}) in Appendix B) by
\begin{eqnarray}
\mu^{2}_{\mathrm{eff}} = \frac{3 m_{2}^{2}}{8\pi (1/2) m_{1}^{2} \Phi_{1}^{2}}.
\end{eqnarray}

The key point is that we have to classicalize not only the $\Phi_{1}$ direction, but also the $\Phi_{2}$ direction. For the single field case, around the local minimum, the classicality condition is satisfied if $\mu_{\mathrm{eff}} < 3/2$ (see Appendix B, also the analysis of \cite{Hartle:2008ng} supports this bound). We can regard that along the $\Phi_{2}$ direction, we have to check the classicality of the local minimum and hence $\mu_{\mathrm{eff}} < 3/2$ is a good criterion, as long as $\Phi_{1}$ direction is sufficiently gentle. This means that the classicality can be satisfied for the $\Phi_{2}$ direction, if
\begin{eqnarray}
\mu^{2}_{\mathrm{eff}} = \frac{3 m_{2}^{2}}{8\pi (1/2) m_{1}^{2} \Phi_{1}^{2}}< \frac{9}{4},
\end{eqnarray}
or equivalently,
\begin{eqnarray}
\Phi_{1} > \frac{1}{\sqrt{3\pi}} \times \frac{m_{2}}{m_{1}} \simeq 0.34 \times \frac{m_{2}}{m_{1}} \left(\equiv \Phi_{1,\mathrm{m}}\right),
\end{eqnarray}
where $\Phi_{1,\mathrm{m}}$ will do the role of the cutoff along the $\Phi_{1}$ direction.

For this condition, the number of $e$-foldings are \cite{Linde:1983gd}\footnote{Note that study on $e$-foldings of general two field inflation models is discussed in \cite{St}.}
\begin{eqnarray}
\mathcal{N} = 2 \pi \Phi_{1}^{2} > 2 \pi \Phi_{1,\mathrm{m}}^{2} = 0.67 \times \left(\frac{m_{2}}{m_{1}}\right)^{2}.\label{eq:theore}
\end{eqnarray}
This approximation is true for the $m_{1} \ll m_{2}$ limit. We confirm these limiting behaviors by using numerical calculations.

\subsection{Numerical confirmations}

\subsubsection{Shape of cutoffs}

\begin{figure}
\begin{center}
\includegraphics[scale=0.45]{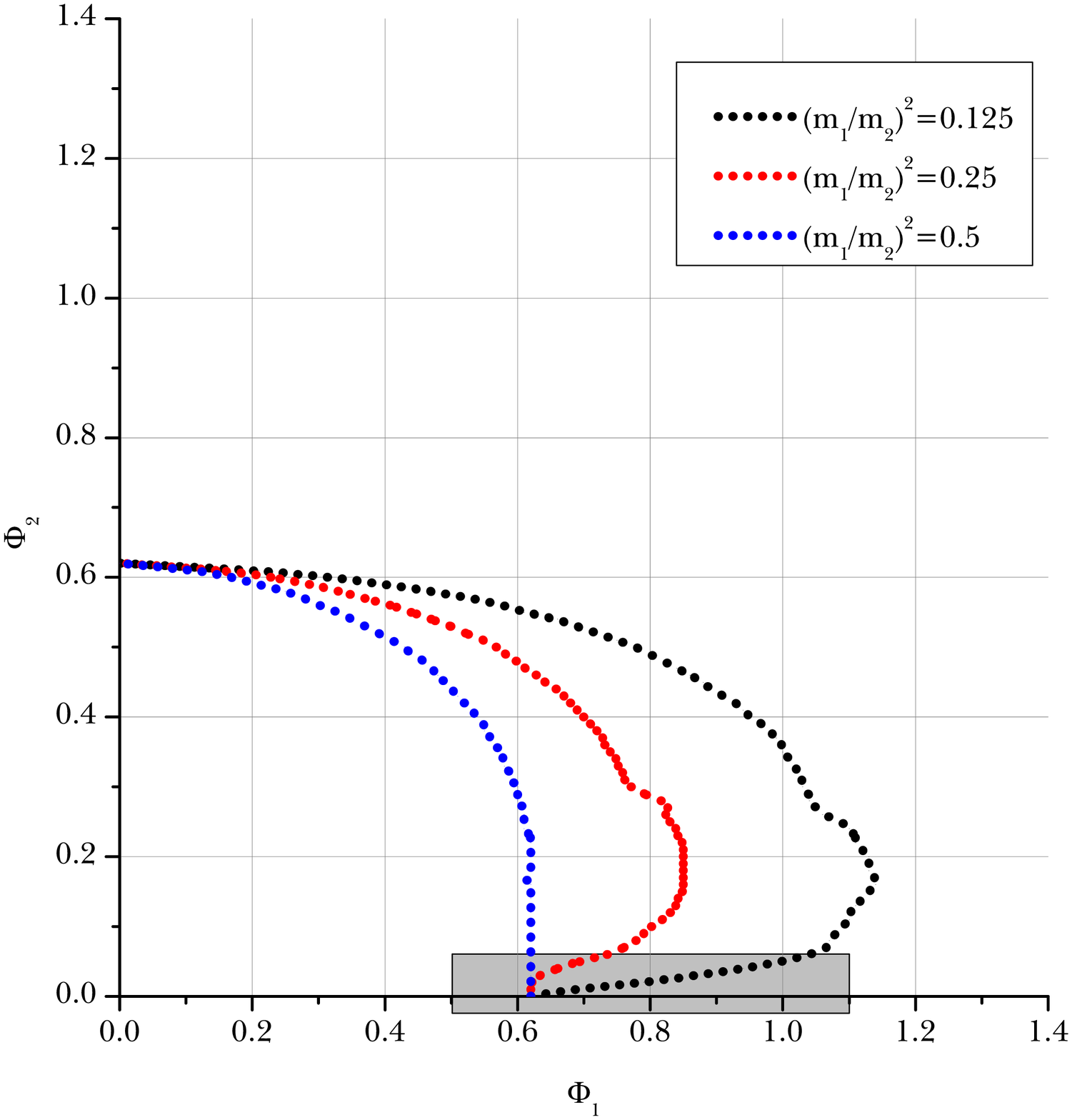}
\includegraphics[scale=0.4]{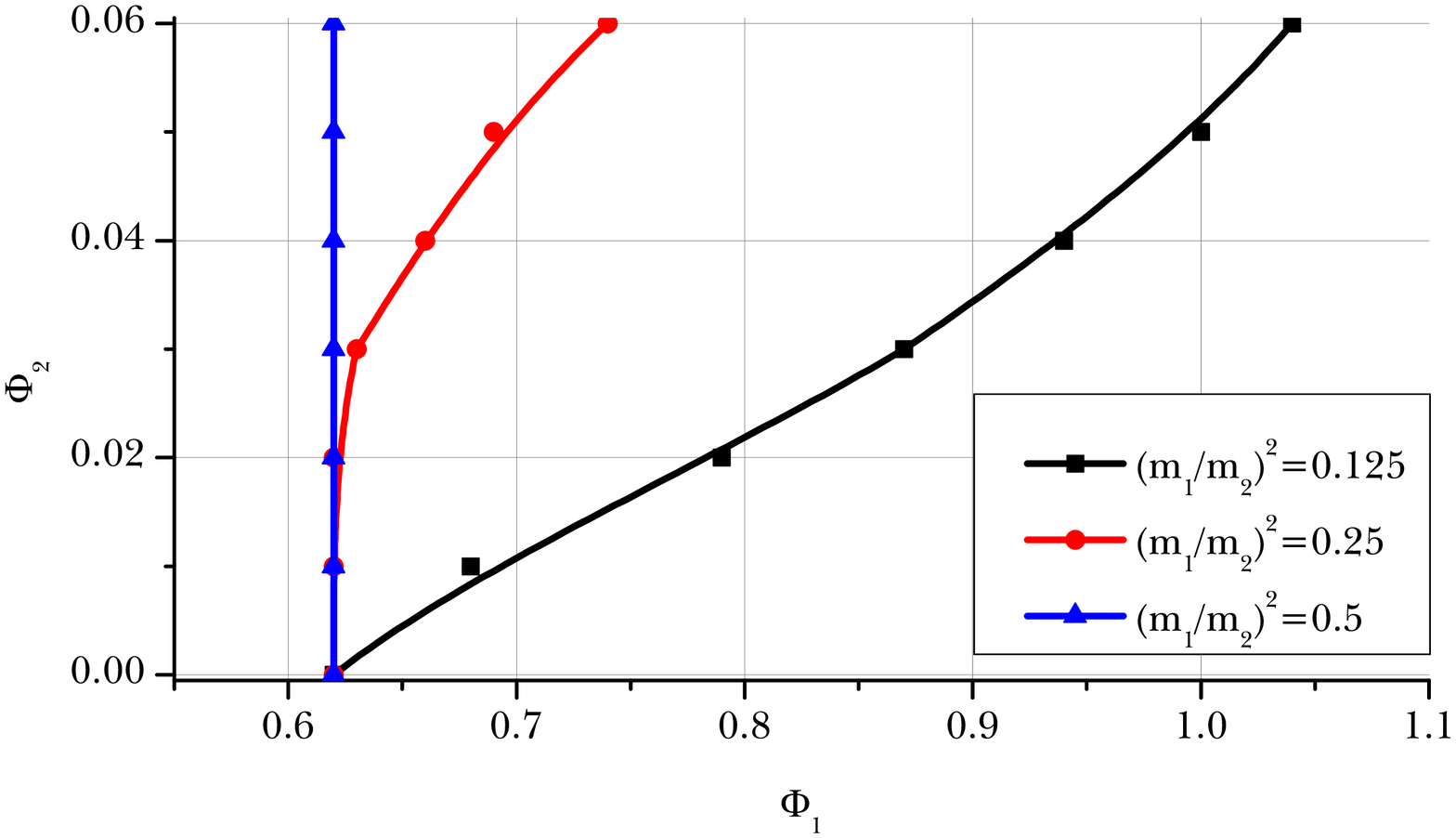}
\caption{\label{fig:cutoffs_new}Upper: Numerical calculations of the cutoffs for $(m_{1}/m_{2})^{2} = 0.125, 0.25$, and $0.5$. As the ratio $m_{1}/m_{2}$ decreases, the cutoff shape is tilted. Lower: The magnification of the upper gray colored box, near the maximum probability point.}
\end{center}
\end{figure}

Upper of Figure~\ref{fig:cutoffs_new} shows the shape of cutoffs as varying the ratio $m_{1}/m_{2}$. Due to the limitation of the numerical searching, the shape is not entirely smooth, but this is enough to show the general behavior.

\begin{figure}
\begin{center}
\includegraphics[scale=0.75]{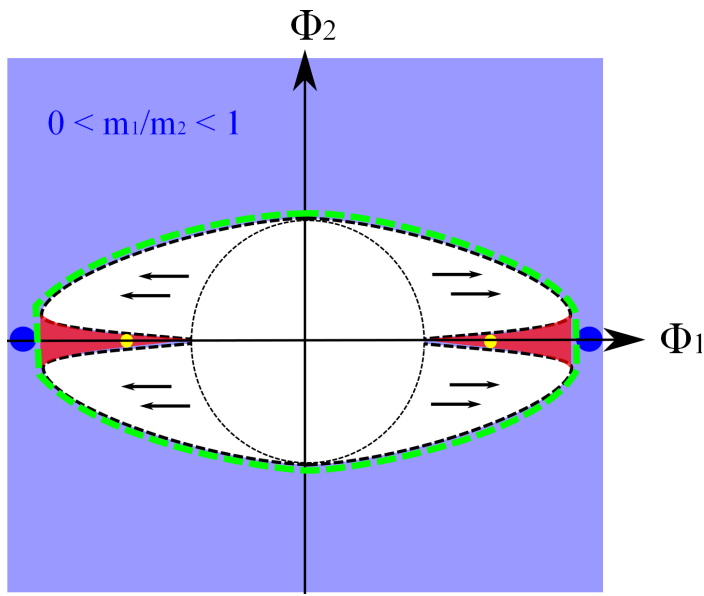}
\caption{\label{fig:concept3}Summary of our results. Black dashed curve is the location of the cutoff. As $m_{1}/m_{2}$ decreases, the cutoff is tilted as the black arrows direct. The red colored region will be narrower and narrower and hence the most probable points (yellow dots) will be negligible. Then, the next effectively most probable point will be around the blue dots.}
\end{center}
\end{figure}

As $m_{1}/m_{2}$ decreases, the shape of cutoffs drastically changed. One has to observe by two issues. One is the narrow and sharp region of the slow-direction (red colored region in Figure~\ref{fig:concept3}). The other is the wider and approximately elliptic region (green dashed curve in Figure~\ref{fig:concept3}).

\paragraph{Near the slow direction} In lower of Figure~\ref{fig:cutoffs_new}, we show that near the cutoff, for a given $\Phi_{1}$, the classicalizable range of $\Delta \Phi_{2}$ becomes narrower and narrower. Therefore, it is reasonable to see that as $m_{1}/m_{2} \rightarrow 0$, the range $\Delta \Phi_{2}$ will approach zero and hence we will restore the lower of Figure~\ref{fig:concept2} and such a narrow region will emerge to one dimensional slices $\Phi_{2} = 0$ and $|\Phi_{1}| > \Phi_{c}$.

\begin{figure}
\begin{center}
\includegraphics[scale=0.4]{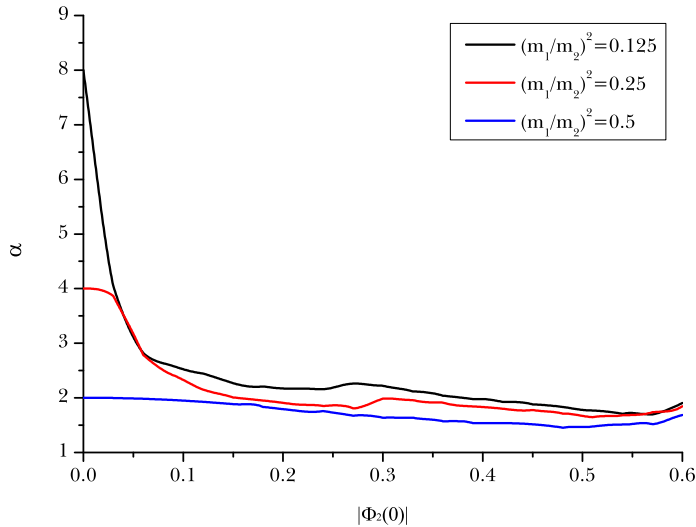}
\caption{\label{fig:alpha}Plot of $\left(\frac{m_{2}}{m_{1}}\right)^{2} \frac{\Phi_{c}^{2} - \Phi_{2}^{2}}{\Phi_{1}^{2}}$ through the cutoff. This approaches an order one parameter as $\Phi_{2}$ increases. Therefore, Equation~(\ref{eq:fit}) is a good approximation to describe the green dashed curve of Figure~\ref{fig:concept3}.}
\end{center}
\end{figure}

\paragraph{Boundary of cutoffs} As we see in the Figure~\ref{fig:alpha}, we tried to fit the cutoff by the following function:
\begin{eqnarray}\label{eq:fit}
\Phi_{1}^{2} = \left(\frac{m_{2}}{m_{1}}\right)^{2} \frac{\Phi_{c}^{2} - \Phi_{2}^{2}}{\alpha},
\end{eqnarray}
where $\Phi_{c} \simeq 0.62$ is the location of the cutoff and $\alpha$ is the fitting parameter. We can observe that $\alpha \simeq \mathcal{O}(1)$ is an order one constant for $|\Phi_{2}(0)| \gtrsim 0.2$. Therefore, this is enough to fit the boundary of the cutoffs (the green dashed curve of Figure~\ref{fig:concept3}).

\subsubsection{Probability distribution}

\begin{figure}
\begin{center}
\includegraphics[scale=0.25]{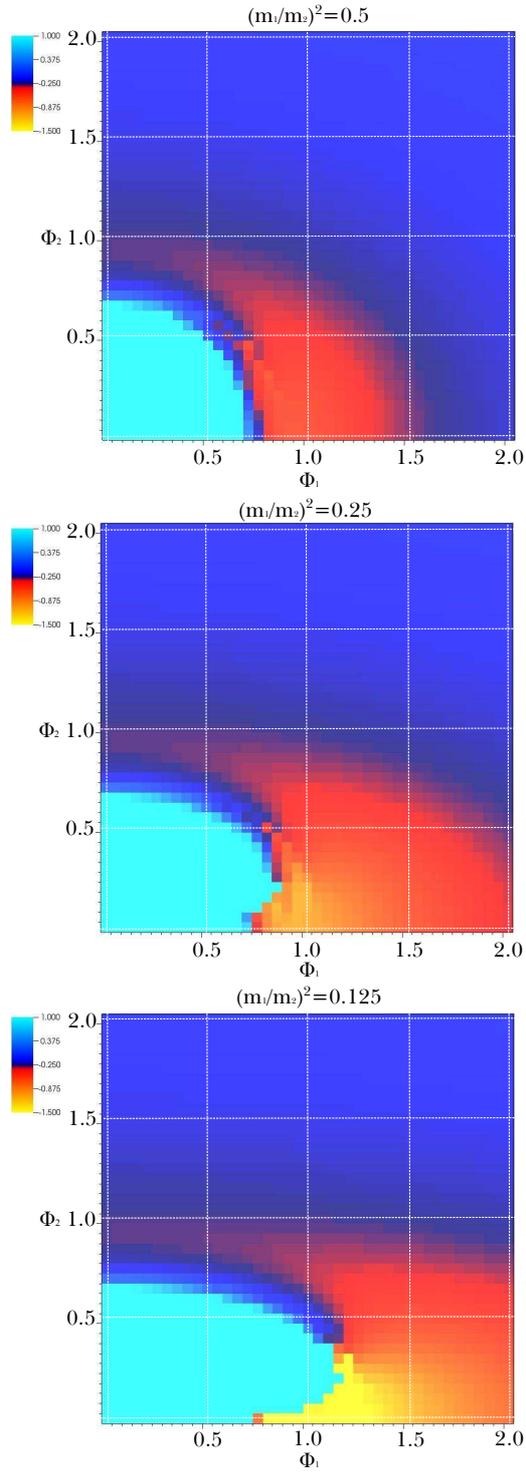}
\caption{\label{fig:ACTION}Euclidean action by varying $m_{1}/m_{2}$. Sky-blue region corresponds either inside the cutoff or has the Euclidean action more than one (hence highly suppressed).}
\end{center}
\end{figure}

Figure~\ref{fig:ACTION} gives the plots of Euclidean actions as we vary $m_{1}/m_{2}$. By varying $|\Phi_{1}(0)|$ and $|\Phi_{2}(0)|$, we found the classicalized instantons and plotted the Euclidean actions. We searched $0 \leq \Phi_{1} \leq 2$ and $0 \leq \Phi_{2} \leq 2$ with the step size $0.5$, and hence the number of simulation points are $41 \times 41 = 1681$ for each figure.

As we see in this figure, the most probable point is near $\Phi_{1} \simeq 1$ and $\Phi_{2} = 0$, i.e., it is located in the $\Phi_{1}$ axis. However, as we observed in Figure~\ref{fig:cutoffs_new}, as $m_{1}/m_{2}$ decreases, the field space near the most probable point will be narrower and narrower and in the end the point will be negligible. Then, as $m_{1}/m_{2} \ll 1$, the realistic physical cutoff will emerge to the green dashed curve in Figure~\ref{fig:concept3}. Along the green dashed curve, the most probable point is on the line $\Phi_{2} = 0$ (the blue dots in Figure~\ref{fig:concept3}).

\subsubsection{Implications for inflation}

The cutoff is a circular shape $|\Phi| = \Phi_{c}$ for $m_{1}=m_{2}$. As $m_{1}/m_{2}$ decreases, the cutoff is tilted and the unclassicalized region increases along the slow direction. Through the $\Phi_{2} = 0$ line, the classicalizable field space will be narrower and narrower, and hence eventually, it will be negligible.

Then Equation~(\ref{eq:fit}) will be the effective cutoff (green dashed curve in Figure~\ref{fig:concept3}). On this effective cutoff, the (effectively) most probable point is on the line $\Phi_{2} = 0$. Then, the following relation is satisfied:
\begin{eqnarray}\label{eq:rel}
V(\Phi_{1,\mathrm{m}},\Phi_{2}=0) = \frac{m_{1}^{2}}{2} \Phi_{1,\mathrm{m}}^{2} = \frac{m_{2}^{2}}{2\alpha} \Phi_{c}^{2},
\end{eqnarray}
where $\Phi_{1,\mathrm{m}}$ denotes the point that has the maximum probability on the cutoff surface Equation~(\ref{eq:fit}). For this point, the expected number of $e$-foldings are \cite{Linde:1983gd}
\begin{eqnarray}\label{eq:efold}
\mathcal{N} = 2 \pi \Phi_{1,\mathrm{m}}^{2} \simeq \frac{2.5}{\alpha} \times \left(\frac{m_{2}}{m_{1}}\right)^{2}.
\end{eqnarray}
If we compare with the previous result of theoretical expectations in Equation~(\ref{eq:theore}), $\alpha \sim 3.3$ is obtained as an order one parameter, and hence this is consistent with Figure~\ref{fig:alpha}.

This means that if we require
\begin{itemize}
\item[(1)] the classicalization of all fields,
\item[(2)] the existence of massive field direction with sufficient mass hierarchy,
\end{itemize}
then these enhance larger $e$-foldings compared to the single mass case.

\subsection{Summary}

\subsubsection{Results}

Here, it will be convenient to summarize our results (Figure~\ref{fig:concept3}).
\begin{description}
\item[1. Shape of cutoffs:] For $0 < m_{1}/m_{2} < 1$, the shape of the cutoff changes as noted in Figure~\ref{fig:concept3}. As $m_{1}/m_{2}$ decreases from one, the shape changes from the circle to the black dashed curve and tilts through the black arrows.
\item[2. Approach to massless limit:] If $m_{1}/m_{2} \ll 1$, then the red colored region becomes narrower and narrower. In addition, the black dashed curve becomes wider and wider as the black arrow directed. The outer part of the black dashed curve will approach the lines $|\Phi_{2}| = \Phi_{c}$ and the red colored region will approach $\Phi_{2} = 0$ and $|\Phi_{1}|>\Phi_{c}$, where two regions will be separated. In the end, it will approach the lower of Figure~\ref{fig:concept2}.
\item[3. Separation near the slow direction:] The most probable point is on the slow mass direction near the cutoff: yellow dots in Figure~\ref{fig:concept3}. However, since the red colored region will be narrower and narrower, in the end, the contribution for the red colored region and yellow dots will be negligible.
\item[4. Probable initial condition:] If we ignore the red colored region, then the next most effectively probable point is around the blue dotted region. So, if two masses have a large hierarchy, then probably the universe will begin around the blue dots.
\end{description}

\subsubsection{Comments on classicalization}

Finally we comment on two questions for the classicalization of two fields.

First, for the $m_{1}/m_{2} \ll 1$ case, anyway there exist regions that have more probabilities and less $e$-foldings (blue lines in lower of Figure~\ref{fig:concept2}). Then why can we ignore such a region? Regarding this question, this is similar with the single field case. For a single field case with $\mu > 3/2$, there exists a cutoff where the classicality is not allowed if the initial condition is inside the cutoff. However, one important observation is that $\Phi = 0$ (i.e., the initial condition is at the local minimum of the potential) gives an exact solution (a de Sitter space or a Minkowski space) with the largest probability. We should not consider the point in a probability distribution \cite{Hartle:2008ng}, since the parameter space volume of $\Phi=0$ is negligible compared to the outer region of the cutoff. We can check this by using a perturbative analysis (see \cite{Hartle:2008ng} as well as Appendix B of this paper); as long as the initial condition is slightly biased from the exact solution $\Phi = 0$, the history cannot be classicalized. Such thing also happens around the blue lines of Figure~\ref{fig:concept2}. So, as long as $m_{1}/m_{2} \ll 1$ happens, it will be justified that such a small striped region should be ignored in the entire wave function. In general, for an $n$-field system, as long as there are mass hierarchies, we expect the same thing should happen. To classicalize the most massive field, the other fields should be excited and hence should start from large field values. This can help to explain a large number of $e$-foldings.

Second, in the usual way of field theoretical treatments, if $m_{1}/m_{2} \ll 1$, then we ignore the field direction $\Phi_{2}$ and assume that $\Phi_{2} \simeq 0$. However, in our treatments, as $m_{2}$ increases, the importance of the $\Phi_{2}$ direction increases. Is it reasonable? Regarding this question, if we restrict our description in Lorentzian dynamics, of course it is true and we should regard that $\Phi_{2} \simeq 0$. However, if we consider Euclidean dynamics, the larger mass direction is more important, since the Euclidean dynamics is governed not by $V$ but by $-V$. Hence, as $m_{2}$ increases, $\Phi_{2}$ direction becomes more unstable; then, to classicalize $\Phi_{2}$ direction, the initial condition of $\Phi_{1}$ should increase further. This is the reason why such a counter-intuitive thing happens in two-field instantons.

\section{\label{sec:con}Conclusion}

In this paper, we investigated the no-boundary wave function and the complex-valued instantons for two massive field models. If there is a relatively massive direction, then the field is highly unstable along the massive direction in the Euclidean time signature. To classicalize along the massive direction, the instanton should begin from relatively larger vacuum energy. Therefore, the existence of the massive direction implies the increase of $e$-foldings. The expected $e$-foldings are $\mathcal{N} \simeq (m_{2}/m_{1})^{2} \times \mathcal{O}(1)$. As long as there is a mass hierarchy, the no-boundary wave function can reasonably explain sufficient inflation, even more than $50$ $e$-foldings.

The existence of massive direction, in other words, the existence of mass hierarchy can help to explain the large number of $e$-foldings. However, this necessarily requires the super-Planckian field: $\Phi_{1,\mathrm{m}} \sim m_{2}/m_{1} \gg 1$. This potentially unnatural initial condition can be explained by introducing multi-field inflation. If there are $n$ number of scalar fields $\phi_{1}^{i}$ ($i=1,...,n$) with mass $m_{1}$ and one scalar field $\phi_{2}$ with mass $m_{2} \gg m_{1}$, then the most probable initial condition will correspond approximately \cite{Hwang:2012bd}
\begin{eqnarray}
\phi_{1}^{i} \simeq \frac{m_{2}}{m_{1}} \frac{1}{\sqrt{n}}.
\end{eqnarray}
Therefore, the super-Planckian problem can be approximately resolved by introducing a number of fields, though this should be confirmed by more definite calculations.

It is also easy to generalize for multiple mass cases. If there are sufficient hierarchies for different masses, then \textit{the most probable initial condition is the slowest direction with the field value that satisfies all the massive fields (especially, the most massive field) to be classicalized}. Our investigation can be generalized, not only for quadratic fields, but also for different kinds of potentials, e.g., axion type fields, and we remain this for a future work.

The existence of massive field naturally explains the traditional problem of inflation \cite{Ijjas:2013vea} and Euclidean quantum cosmology \cite{Hawking:1984hk}. Although this is not the unique resolution of the problem, anyway now we can say that the no-boundary wave function can explain sufficient $e$-foldings with very conservative assumptions. This may expect some observational implications; e.g., expect the existence of one more massive field that is different from the inflaton field. In principle, this opens a good way to confirm or falsify some expectations that is originated from quantum cosmology.

\section*{Acknowledgment}

We would like to thank to Bum-Hoon Lee and also thank to computer facilities in Center for Quantum Spacetime, Sogang University. DY was supported by the JSPS Grant-in-Aid for Scientific Research (A) No. 21244033 and by Leung Center for Cosmology and Particle Astrophysics (LeCosPA) of National Taiwan University (103R4000). SAK was supported by the Research fund No. 1-2010-2469-001-4 by Ewha Womans University.

\newpage

\section*{Appendix A: Numerical searching algorithm}

In this paper, we used one numerical algorithm to find complex-valued instantons. We already used this technique in the previous paper \cite{Hwang:2013nja,Hwang:2012bd,Hwang:2011mp,Hwang:2012mf} and we need to generalize for multi-field cases. For two scalar field cases, we have twelve initial conditions: the real part and imaginary part of $a(0)$, $\Phi_{1}(0)$, $\Phi_{2}(0)$, $\dot{a}(0)$, $\dot{\Phi}_{1}(0)$, and $\dot{\Phi}_{2}(0)$. Among these conditions, we already fix eight of them, since we require the regularity of $\tau=0$: $a(0)=0$, $\dot{a}(0)=1$, $\dot{\Phi}_{1}(0)=0$, and $\dot{\Phi}_{2}(0)=0$. Now, there remain four initial conditions: $\Phi_{1}(0)=|\Phi_{1}(0)| e^{i\theta_{1}}$ and $\Phi_{2}(0)=|\Phi_{2}(0)| e^{i\theta_{2}}$, where $|\Phi_{1}(0)|$ and $|\Phi_{2}(0)|$ are the modulus of the initial field positions and $\theta_{1}$ and $\theta_{2}$ are the phase angles. In addition, we have to choose a turning point $X$ from the Euclidean time $\tau$ to the Lorentzian time $t$. Therefore, for a complex-valued instanton with a given two field modulus $|\Phi_{1}(0)|$ and $|\Phi_{2}(0)|$, we still have undefined three-dimensional degrees of freedom: $(\theta_{1}, \theta_{2}, X)$. These three parameters should be used to control imaginary parts of $a$, $\Phi_{1}(0)$, and $\Phi_{2}(0)$.

To search proper $(\theta_{1}, \theta_{2}, X)$ to satisfy the classicality condition, we need to use a searching algorithm (so-called the genetic algorithm, where this was discussed in detail by \cite{Hwang:2011mp}). For technical convenience, we minimize the following objective function
\begin{eqnarray}
F_{\phi_{0}} \left[\theta_{1}, \theta_{2}, X \right] \equiv \int_{T_{1}}^{T_{2}} \left| \frac{\mathcal{L}^{\mathrm{Re}}_{\phi_{0}}[\theta_{1},\theta_{2}, X](t)}{ \mathcal{L}_{\phi_{0}}[\theta_{1},\theta_{2}, X](t)}\right| dt.
\end{eqnarray}
Here, $\mathcal{L}_{\phi_{0}}[\theta_{1},\theta_{2}, X](t)$ is the Lagrangian with a given initial field modulus $|\Phi_{1}(0)|$ and $|\Phi_{1}(0)|$. $T_{1}$ and $T_{2}$ are sufficiently large time values when we measure the Lagrangian and are introduced for technical conveniences. After we find a candidate $(\theta_{1}, \theta_{2}, X)$ that minimizes the objective function, we check the classicality again for sure: $|\Phi_{1}^{\mathrm{Im}}/\Phi_{1}^{\mathrm{Re}}| \ll 1$, $|\Phi_{2}^{\mathrm{Im}}/\Phi_{2}^{\mathrm{Re}}| \ll 1$, and $|a^{\mathrm{Im}}/a^{\mathrm{Re}}| \ll 1$.

\section*{Appendix B: Classicality and reality for single field inflation}

For the single field inflation model
\begin{eqnarray}
S_{\text{E}} = - \int d^{4}x \sqrt{+g} \left[ \frac{1}{16\pi} \left(R - 2\Lambda \right) -  \frac{1}{2} (\nabla \Phi)^{2} - V(\Phi) \right],
\end{eqnarray}
it is easier to consider classicality in the perturbative regime. We consider the scalar fields with the quadratic potentials with mass $m$:
\begin{eqnarray}
V(\Phi) = \frac{1}{2} m^{2} \Phi^{2}.
\end{eqnarray}
By defining the metric and functions
\begin{eqnarray}
ds^{2} &=& \frac{3}{\Lambda} \left[N(\lambda)^{2} d\lambda^{2} + a(\lambda)^{2} d\Omega_{3}^{2}\right],\\
\phi &\equiv& \sqrt{\frac{4\pi}{3}} \Phi,\\
\mu &\equiv& \sqrt{\frac{3}{\Lambda}} m,\label{eq:mu}
\end{eqnarray}
we obtain the Euclidean action by
\begin{eqnarray}
S_{\text{E}} = \frac{9 \pi}{4\Lambda} \int N d\lambda \left\{- a \left(\frac{d a}{N d\lambda}\right)^{2} - a +a^{3} + a^{3}\left[\left(\frac{d \phi}{N d\lambda}\right)^{2} + \mu^{2}\phi^{2} \right] \right\}.
\end{eqnarray}
The equations of motions are
\begin{eqnarray}
0 &=& \ddot{a} + a + a \left(2 \dot{\phi}^{2} + \mu^{2} \phi^{2} \right),\\
0 &=& \ddot{\phi} + 3 \frac{\dot{a}}{a} \dot{\phi} - \mu^{2} \phi.
\end{eqnarray}

We assume the case that the background metric is already classicalized, but there is a small field perturbation along the real and imaginary directions. Therefore, we assume
\begin{eqnarray}
a^{\text{Im}} \ll a^{\text{Re}}, \qquad \dot{\phi}^{2} \ll 1, \qquad \mu^{2} \phi^{2} \ll 1,
\end{eqnarray}
and $a^{\text{Re}} \simeq b$ and $\phi^{\text{Re}} \simeq \chi$. In this limit, the complexified Einstein equations in the Lorentzian direction become
\begin{eqnarray}
\ddot{a}^{\text{Re}} - a^{\text{Re}} &\simeq&  0,
\label{Einstein_Eq_a_Re}\\
\ddot{a}^{\text{Im}} - a^{\text{Im}} + 2 a^{\text{Re}} \left[2 \dot{\phi}^{\text{Re}} \dot{\phi}^{\text{Im}} - \mu^{2} \phi^{\text{Re}} \phi^{\text{Im}}\right] &\simeq&  0,
\label{Einstein_Eq_a_Im}\\
\ddot{\phi}^{\text{Re}} + 3 \frac{\dot{a}^{\text{Re}}}{a^{\text{Re}}} \dot{\phi}^{\text{Re}} + \mu^{2} \phi^{\text{Re}} &\simeq& 0,
\label{Einstein_Eq_phi_Re}\\
\ddot{\phi}^{\text{Im}} + 3 \frac{\dot{a}^{\text{Re}}}{a^{\text{Re}}} \dot{\phi}^{\text{Im}} + \mu^{2} \phi^{\text{Im}} &\simeq& 0.
\label{Einstein_Eq_phi_Im}
\end{eqnarray}
Then, the on shell Euclidean action becomes
\begin{align}
S_{\text{E}}^{\text{Re}}(y) &\simeq S_{\text{E}}^{\text{Re}}(\tilde{y}) + \frac{9 \pi}{2 \Lambda} \int_{\tilde{y}}^{y} d y' \Bigl\{- 3 a^{\text{Im}} (a^{\text{Re}})^{2} - \mu^{2} \left[3 a^{\text{Im}} (a^{\text{Re}})^{2} \left((\phi^{\text{Re}})^{2} - (\phi^{\text{Im}})^{2}\right) + 2 (a^{\text{Re}})^{3} \phi^{\text{Re}} \phi^{\text{Im}}\right]\Bigr\},\\
S_{\text{E}}^{\text{Im}}(y) &\simeq S_{\text{E}}^{\text{Im}}(\tilde{y}) + \frac{9 \pi}{2 \Lambda} \int_{\tilde{y}}^{y} d y' \Bigl\{(a^{\text{Re}})^{3} + \mu^{2} \left[(a^{\text{Re}})^{3} \left((\phi^{\text{Re}})^{2} - (\phi^{\text{Im}})^{2}\right) - 6 (a^{\text{Re}})^{2} a^{\text{Im}} \phi^{\text{Re}} \phi^{\text{Im}}\right]\Bigr\},
\end{align}
where $\tilde{y}$ denotes any moment satisfying above assumptions and $\int dy$ can be approximated by a small time step $\Delta y$. Therefore, approximately,
\begin{align}
|\nabla_{b} S_{\text{E}}^{\text{Re}}[b,\chi]| &\simeq \frac{9 \pi}{2 \Lambda} \Delta y \left|- 6 a^{\text{Im}} a^{\text{Re}} - \mu^{2} \left[6 a^{\text{Im}} a^{\text{Re}} \left((\phi^{\text{Re}})^{2} - (\phi^{\text{Im}})^{2}\right) + 6 (a^{\text{Re}})^{2} \phi^{\text{Re}} \phi^{\text{Im}}\right]\right|\\
 &\simeq \frac{9 \pi}{2 \Lambda} \Delta y \left| 6 \phi^{\text{Re}} \phi^{\text{Im}} \right| (a^{\text{Re}})^{2},\\
|\nabla_{b} S_{\text{E}}^{\text{Im}}[b,\chi]| &\simeq \frac{9 \pi}{2 \Lambda} \Delta y \left|3 (a^{\text{Re}})^{2} + \mu^{2} \left[3 (a^{\text{Re}})^{2} \left((\phi^{\text{Re}})^{2} - (\phi^{\text{Im}})^{2}\right) - 12 a^{\text{Re}} a^{\text{Im}} \phi^{\text{Re}} \phi^{\text{Im}}\right]\right|\\
&\simeq \frac{9 \pi}{2 \Lambda} \Delta y \left|3 + 3 \mu^{2} \left((\phi^{\text{Re}})^{2} - (\phi^{\text{Im}})^{2}\right)\right| (a^{\text{Re}})^{2},\\
|\nabla_{\chi} S_{\text{E}}^{\text{Re}}[b,\chi]| &\simeq  \frac{9 \pi}{2 \Lambda} \Delta y \left|- \mu^{2} \left[6 a^{\text{Im}} (a^{\text{Re}})^{2} \phi^{\text{Re}} + 2 (a^{\text{Re}})^{3} \phi^{\text{Im}}\right]\right|\\
&\simeq  \frac{9 \pi}{2 \Lambda} \Delta y \mu^{2} \left|2 \phi^{\text{Im}}\right| (a^{\text{Re}})^{3},\\
|\nabla_{\chi} S_{\text{E}}^{\text{Im}}[b,\chi]| &\simeq  \frac{9 \pi}{2 \Lambda} \Delta y \left|\mu^{2} \left[2(a^{\text{Re}})^{3} \phi^{\text{Re}} - 6 (a^{\text{Re}})^{2} a^{\text{Im}} \phi^{\text{Im}}\right]\right|\\
&\simeq  \frac{9 \pi}{2 \Lambda} \Delta y \mu^{2} \left| 2 \phi^{\text{Re}} \right|(a^{\text{Re}})^{3}.
\end{align}
In this limit, the classicality ratios
\begin{eqnarray}
Cl_{b} &\equiv& \frac{|\nabla_{b} S_{\text{E}}^{\text{Re}}[b,\chi]|}{|\nabla_{b} S_{\text{E}}^{\text{Im}}[b,\chi]|},\\
Cl_{\chi} &\equiv& \frac{|\nabla_{\chi} S_{\text{E}}^{\text{Re}}[b,\chi]|}{|\nabla_{\chi} S_{\text{E}}^{\text{Im}}[b,\chi]|} \simeq \frac{\phi^{\text{Im}}}{\phi^{\text{Re}}}
\end{eqnarray}
become reasonably small if and only if $\phi^{\text{Im}} \ll \phi^{\text{Re}}$ is satisfied. Therefore, it is reasonable to understand that the classicality conditions are equivalent to the following conditions:
\begin{eqnarray}
a^{\text{Im}} \ll a^{\text{Re}}, \quad \phi^{\text{Im}} \ll \phi^{\text{Re}}.
\end{eqnarray}

We already assumed $a^{\text{Im}} \ll a^{\text{Re}}$. Then, the question is on the scalar field. \textit{For which condition of the potential, can the classicality be satisfied?}

\begin{description}
\item[-- Case 1: $\mu < 3/2$:]
We first consider a large cosmological constant, assuming $\mu < 3/2$. Then, Equation~\eqref{Einstein_Eq_a_Re} has a general solution
\begin{eqnarray}
a^{\text{Re}} = C_{a}^{\text{Re}} e^{t} + D_{a}^{\text{Re}} e^{-t} \simeq C_{a}^{\text{Re}} e^{t},
\end{eqnarray}
where $C$ and $D$ are integration constants. Then, Equation~\eqref{Einstein_Eq_phi_Re} and \eqref{Einstein_Eq_phi_Im} give
\begin{eqnarray}
\phi^{\text{Re}} &=& C_{\phi}^{\text{Re}} e^{-\frac{3}{2} t + \omega t} + D_{\phi}^{\text{Re}} e^{-\frac{3}{2} t - \omega t},\\
\phi^{\text{Im}} &=& C_{\phi}^{\text{Im}} e^{-\frac{3}{2} t + \omega t} + D_{\phi}^{\text{Im}} e^{-\frac{3}{2} t - \omega t}.
\end{eqnarray}
By tuning the initial condition, one can fix a constant $C_{\phi}^{\text{Im}} \ll 1$. Then, $\phi^{\text{Im}}/\phi^{\text{Re}} \simeq e^{-2\omega t}$ and hence the classicality can be obtained.

\item[-- Case 2: $\mu > 3/2$:]
We second consider a small cosmological constant, assuming $\mu > 3/2$.

We are interested in a classical history which has expanding $a^{\text{Re}}$ and decaying $a^{\text{Im}}$ and $\phi^{\text{Im}}$. The above set of differential equations has a solution satisfying above requirements,
\begin{eqnarray}
\phi^{\text{Re}} &=& C_{\phi}^{\text{Re}} e^{-\frac{3}{2} t} \cos(\omega t + \alpha),\\
\phi^{\text{Im}} &=& C_{\phi}^{\text{Im}} e^{-\frac{3}{2} t} \cos(\omega t + \beta),
\end{eqnarray}
where
\begin{eqnarray}
\omega \equiv \sqrt{\mu^{2} - \left(\frac{3}{2}\right)^{2}}
\end{eqnarray}
and some real constants $C_{\phi}^{\text{Re}}$, $C_{\phi}^{\text{Im}}$, $\alpha$ and $\beta$. In this case, we cannot choose $C_{\phi}^{\text{Im}}=0$, because it makes $\phi^{\text{Im}} = 0$ and hence this cannot be regular at $\tau = 0$ unless $\mu = 0$ (i.e., unless it is the exact Hawking-Moss instanton \cite{Hawking:1981fz}, the scalar field should have non-vanishing imaginary part due to the analyticity). Then, for any initial conditions, $\phi^{\text{Im}}/\phi^{\text{Re}} \simeq O(1)$ and hence the classicality cannot be obtained.
\end{description}

To summarize, this explains why there should exist the cutoff around the local minimum when $\mu > 3/2$. By the same reasoning, for the multi-field inflation, we need to consider the classicality. If one field is slow-roll, then it is rather easy to find the cutoff using the effective mass.

\end{document}